\documentclass[final,twocolumn,12pt]{elsarticle}

\journal{Intermetallics}
\usepackage{graphicx}% Include figure files
\begin{document}

\begin{frontmatter}

\title{Effect of the temperature and magnetic field induced martensitic transformation in bulk
Fe$_{45}$Mn$_{26}$Ga$_{29}$ alloy on its electronic structure and
physical properties}

\author[label1]{Y. V. Kudryavtsev \corref(\fnref{fn1}}

\fntext[fn1]{Corresponding author}

\ead{kudr@imp.kiev.ua}

\author[label1]{N. V. Uvarov}
\author[label1]{A. E. Perekos}
\address[label1]{Institute of Metal Physics, NAS of Ukraine, Vernadsky 36,03142, Kiev, Ukraine}

\author[label2]{J. Dubowik}
\address[label2]{Institute of Molecular Physics, PAS, M. Smoluchowskiego
17, 60-179 Pozna\'{n}, Poland}

\author[label3]{L. E. Kozlova}
\address[label3]{Institute of Magnetism, NAS of Ukraine, Vernadsky
36b, 03142, Kiev, Ukraine}

\begin{abstract}

Effect of the temperature and magnetic field induced martensitic
transformation (MT) on the electronic structure and some physical
properties of bulk Fe$_{45.2}$Mn$_{25.9}$Ga$_{28.9}$ Heusler alloy
has been investigated. {According to the experimental
results of DSC, magnetic and transport measurements direct and
reverse martensitic transformation without external magnetic field
takes place within 194 $\leq T \leq$ 328 K temperature range with
a hysteresis up to $\Delta T \approx$ 100 K defined as $\Delta T$
= $A_{f,s}$ - $M_{s,f}$, where $A_{f,s}$ and $M_{s,f}$ are the
critical temperatures of direct and reverse martensitic
transformation. External magnetic field of $\mu_{0}H$ = 5 T causes
a high-temperature shift of MT temperatures.} MT from parent
austenite L2$_{1}$ phase to martensitic tetragonally distorted
L2$_{1}$ one (i. e. to L1$_{0}$) causes significant changes in the
electronic structure of alloy, a drastic increase in alloy
magnetization, a decrease in the alloy resistivity, and a reversal
of sign of the temperature coefficient of resistivity from
negative to positive. At the same time experimentally determined
optical properties of Fe$_{45.6}$Mn$_{25.9}$Ga$_{28.9}$ Heusler
alloy in austenitic and martensitic states look visually rather
similar being noticeable different in microscopic nature as can be
concluded from first-principle calculations. Experimentally
observed changes in the physical properties of the alloy are
discussed in terms of the electronic structures of an austenite
and martensite phases.

\end{abstract}


\end{frontmatter}

\section{Introduction}
\label{1}

Magnetic shape memory alloys (MSMAs) have found significant
attention due to the possibility of rearrangement of martensite
variants by external magnetic field. This property opens wide
perspectives of their practical applications in medicine,
robotics, active or passive damping \cite{jenkins}. Among MSMAs
Ni-based Heusler alloys (HA) like Ni$_{2}$MnGa, Ni$_{2}$FeGa or
Ni$_{2}$MnAl probably are most investigated
\cite{ni2mnga,ni2fega,ni2mnal}. In Ni-based MSMA HAs direct
martensitic transformation (MT) from ferromagnetic (FM) austenite
phase to FM martensite one is accompanied with a small reduction
in alloy magnetization and an increase in alloy resistivity.

Quite distinct changes in these properties induced by a direct MT
have been found in off-stoichiometric Fe$_{2}$MnGa alloys
\cite{zhu,omori1,omori2,hovaj}. The direct MT in these alloys is
accompanied with a significant decrease in alloy resistivity and a
transition from paramagnetic (PM) austenite to an FM martensite
phase.

Phase equilibria upon phase transitions (like MT) are usually
discussed in the framework of thermodynamics \cite{vasil}. Effect
of the electronic structures on such structural transformations
was taken into account in a few reports \cite{entel1,entel2}.

Among various experimental methods for studying of electronic
structures of metals the optical spectroscopy (namely
spectroscopic ellipsometry) is usually considered as such that
manifests a higher compared to x-ray spectroscopy energy
resolution ($\sim$ 0.01 eV) within $\pm$ 5 - 6 eV energy range
near the Fermi level ($E_{F}$). Hence, spectroscopic ellipsometry
was successfully employed for studying the solid-state reactions
in multilayered films \cite{mlf}, structural transformations in
bulk metals and alloys \cite{structrans}. For example, Sasovskaya
\emph{et al.} have shown experimentally that austenite to
martensite transformation in NiTi alloy is accompanied with the
appearance of a new intense absorption band in a near infrared
region of spectra for martensite state \cite{saso}. It was also
theoretically predicted that MT in Ni$_{2}$MnGa HA results from
changes the electronic structure of alloy as well as its optical
properties \cite{wan}. {Furthermore, Gan'shina \emph{et
al.} have shown that the temperature induced martensitic
transformation in Fe$_{48}$Mn$_{24}$Ga$_{28}$ alloy causes visible
changes in the equatorial Kerr effect spectra which reflect the
changes in the electronic structure and magnetic properties of
alloy \cite{ganshi}.} In this work, we try to consider the effect
of the temperature and magnetic field induced MT in Fe$_{2}$MnGa
alloy on its physical properties in close relation to the changes
in the electronic structures of alloy studied theoretically and
experimentally by employing the spectroscopic ellipsometry.

\section{Experimental details}
\label{2}

A slightly off-stoichiometric bulk polycrystalline Fe$_{2}$MnGa
alloy was prepared by melting together pieces of Fe, Mn, and Ga of
99.99\% purity in an arc furnace with a water-cooled Cu hearth
under a 1.3 bar Ar atmosphere. The Ar gas in the furnace before
melting was additionally purified by multiple remelting of a
Ti$_{50}$Zr$_{50}$ alloy getter. To promote the volume
homogeneity, the ingots were remelted five times. After ingot
melting, the weight loss was {about 3 $\%$}.

The actual composition of the fabricated bulk Fe$_{2}$MnGa HA
sample was evaluated by using energy dispersive x-ray spectroscopy
and found to be Fe$_{45.2}$Mn$_{25.9}$Ga$_{28.9}$.

The structural characterization of the sample was carried out at
room temperature (RT) employing x-ray diffraction (XRD) in
${\theta}-$2${\theta}$ geometry with Co-K$_{\alpha }$
(${\lambda}$=0.17902 nm).

An SC 404 F1 Pegasus differential scanning calorimeter (DSC) was
used to determine the phase transformation temperatures.

Magnetic properties of the bulk Fe$_{45.2}$Mn$_{25.9}$Ga$_{28.9}$
alloy were investigated over a temperature range $80\leq T\leq370$
K by measuring the DC-magnetic susceptibility in a weak magnetic
field of 5 Oe and by measuring the magnetization over a
temperature range $4 \leq T \leq 350$ K and a range of magnetic
fields $0.5 \leq H \leq 50 $ kOe by using the PPMS-P7000 system.

Transport properties were measured by using the four-probe
technique over a range of temperatures $80\leq T\leq440$ K using
the sample of $2.20\times1.25\times10.00$ mm$^{3}$ in size.

A bulk sample for optical measurements of about 10 $\times$ 30
$\times$ 2 mm$^{3}$ in size was cut from the ingot employing spark
erosion technique followed by mechanical polishing with diamond
pasts. To eliminate surface contaminations induced by mechanical
polishing, the sample before optical measurements was annealed at
$T$ = 573 K during 150 minutes at high vacuum conditions.

Optical properties [$Re({\sigma}) = \omega\varepsilon_{2}/4\pi$
and $\varepsilon_{1}$, where $\sigma$ is the optical conductivity
(OC), $\varepsilon_{1}$ and $\varepsilon_{2}$ are the real and
imaginary parts of the diagonal components of the dielectric
function
$\widetilde{\varepsilon}=\varepsilon_{1}-i\varepsilon_{2}$] of the
samples were measured by using a spectroscopic rotating-analyzer
ellipsometer in a spectral range of 250 - 2500 nm (5.0 - 0.5 eV)
at a fixed incidence angle of 73$^{\circ}$ at $T$ = 173 and 373 K,
respectively.

\section{Results and discussion}
\label{3}
\subsection{Electronic structure}
\label{3.1}

\begin{figure}[t]
\includegraphics[width=8.5cm,keepaspectratio]{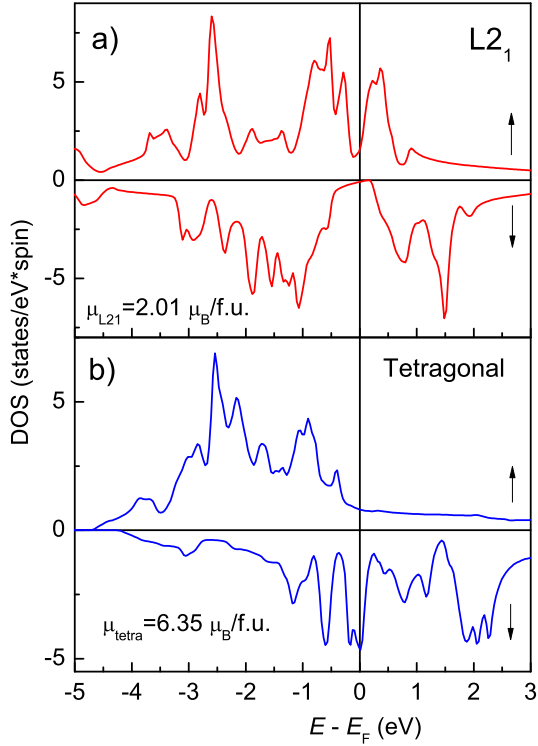}
\caption{Spin-resolved density of states of stoichiometric
Fe$_{2}$MnGa alloy calculated for a) L2$_{1}$ and b) tetragonal
types of structure.} \label{fig1}
\end{figure}

Figure \ref{fig1} shows the spin-resolved energy dependencies of
the density of electronic states [$N(E)$, DOS] for the
stoichiometric Fe$_{2}$MnGa alloy with L2$_{1}$ and tetragonally
distorted L2$_{1}$ (i. e. L1$_{0}$) types of structure calculated
for lattice constants obtained from volume optimization procedure
(for the L2$_{1}$ phase) and from experiment (for tetragonal
phase). Calculations details can be found elsewhere \cite{kudr1}.
It is seen that $N(E)$ dependence for L2$_{1}$ phase is
characterized by energy gap near Fermi level ($E_{F}$) for
minority bands and deep minimum for majority bands making this
phase even not half-metallic but almost semi-metallic. At the same
time for tetragonal phase narrow and intense maximum of the $N(E)$
dependence is observed at $E_{F}$ for minority bands. It can be
expected that such a drastic difference in DOS values at $E_{F}$
for L2$_{1}$ and L1$_{0}$ phases will lead to some difference in
transport properties of alloy and stability of these phases.
Calculated magnetic moments for L2$_{1}$ and L1$_{0}$ phases of
Fe$_{2}$MnGa alloy are ${\mu}=$ 2.01 and 6.35 $\mu_{B}/f.u.$,
respectively.

\subsection{Crystalline structure}
\label{4}

\begin{figure}[t]
\includegraphics[width=8.5cm]{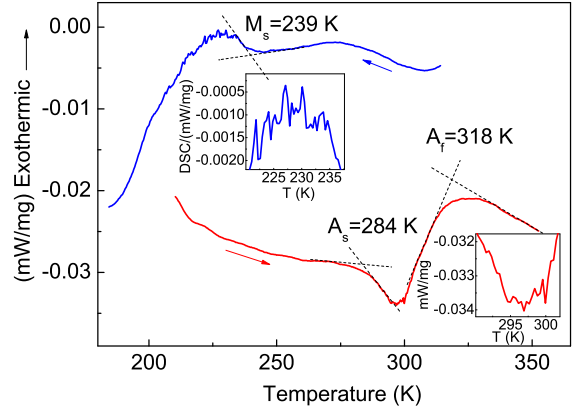}
\caption{DSC patterns of bulk polycrystalline
Fe$_{45.2}$Mn$_{25.9}$Ga$_{28.9}$ alloy during its cooling and
heating. Insets show the enlarged view of the exothermic and
endothermic parts of DSC plots.}\label{fig2}
\end{figure}

\begin{figure}[t]
\includegraphics[width=8.5cm]{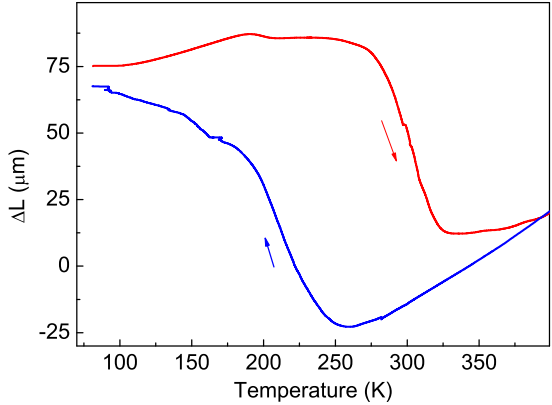}
\caption{Effect of the temperature on extension of bulk
Fe$_{45.2}$Mn$_{25.9}$Ga$_{28.9}$ alloy upon its cooling and
heating.}\label{fig3}
\end{figure}

DSC measurements on sample heating and cooling clearly show an
existence of exo- and endothermic processes displaced in the
temperature and concerned probably with direct and reverse
martensitic transformations in the alloy (see Fig. \ref{fig2})
\cite{zhu,omori1}. Unfortunately, $M_{f}$ temperature using DSC
plot is impossible to determine. Fine structures of DSC plots
within temperature ranges of direct and reverse martensitic
transformations indicate on the multiple-step-transformation
process [i. e. a step-like process of increase (decrease) in a
volume of martensitic phase] in alloy taking place at somewhat
different temperatures due to probably compositional
heterogeneousness. Obtained DSC data agree with the results of
dilatometric measurements of bulk
Fe$_{45.2}$Mn$_{25.9}$Ga$_{28.9}$ alloy upon its cooling and
heating (see Fig. \ref{fig3}). These results show that direct
martensitic transformation is started much below RT at
$M_{s}\approx$ 240 - 255 K, while the reverse martensitic
transformation is finished above RT at $A_{f}\approx$ 318 - 327 K.
Thus, corresponding heat treatments will allow us to fix at RT
either austenite or martensitic phases.

\begin{figure}[t]
\includegraphics[width=8.5cm]{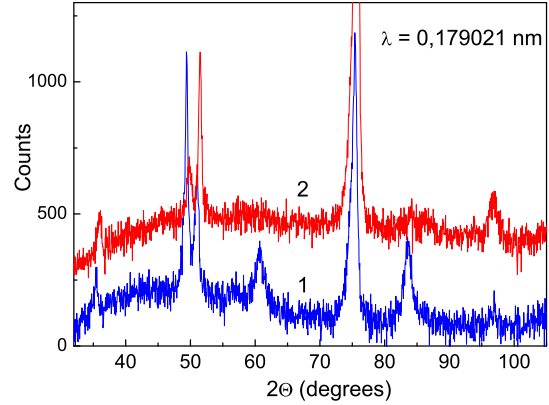}
\caption{RT experimental XRD patterns for bulk
Fe$_{45.2}$Mn$_{25.9}$Ga$_{28.9}$ alloy sample subjected to 1) its
cooling down to $T$ = 78 K and 2) heating up to $T$ = 373 K.}
\label{fig4}
\end{figure}

\begin{figure}[t]
\includegraphics[width=8.5cm]{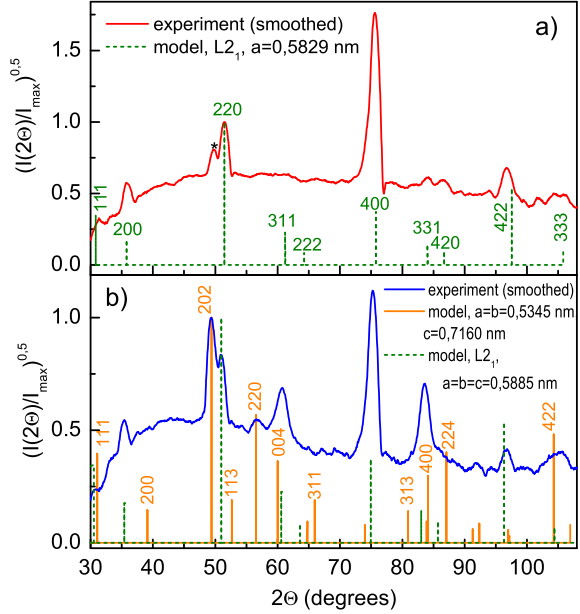}\caption{Smoothed RT
 experimental XRD patterns for bulk
Fe$_{45.2}$Mn$_{25.9}$Ga$_{28.9}$ alloy sample after: a) its
heating up to $T$ = 373 K (red line) and b) after cooling down to
$T$ = 78 K (blue line) together with calculated stroke-diagrams
for stoichiometric Fe$_{2}$MnGa alloy with L2$_{1}$ (green lines)
and tetragonal (yellow line) types of lattice.} \label{fig5}
\end{figure}

Figure \ref{fig4} presents RT experimental XRD patterns of
preliminary heat treated Fe$_{45.2}$Mn$_{25.9}$Ga$_{28.9}$ alloy
sample. The comparison of the obtained experimental XRD patterns
with the results of XRD modelling (see Fig. \ref{fig5}) allows us
to conclude that after heating up to $T$=373 K
Fe$_{45.2}$Mn$_{25.9}$Ga$_{28.9}$ HA at RT contains mainly
L2$_{1}$ phase with the lattice parameters of $a=b=c$ = 0.5829 nm
[superstructure lines (111), (200), (311), (331) and (420) are
clearly seen] with the small admixture of, probably, tetragonal
phase (reflection at 2$\Theta$ = 49.77$^{\circ}$). After cooling
down to $T$ = 78 K at RT the Fe$_{45.2}$Mn$_{25.9}$Ga$_{28.9}$
alloy sample contains two phases. These are mainly body-centered
tetragonal (tetragonally distorted L2$_{1}$) phase with the
lattice parameters of $a=b=0.5345$ nm, $c=0.716$ nm and some
amount of L2$_{1}$ phase with $a=b=c=0.5885$ nm (see Fig.
\ref{fig5}). Even though superstructure reflections for L2$_{1}$
phase at this pattern are hard to be seen it is unlikely that
cooling down to $T=$ 78 K causes significant atomic disorder in
this phase of alloy.

\begin{figure}[t]
\includegraphics[width=6.9cm]{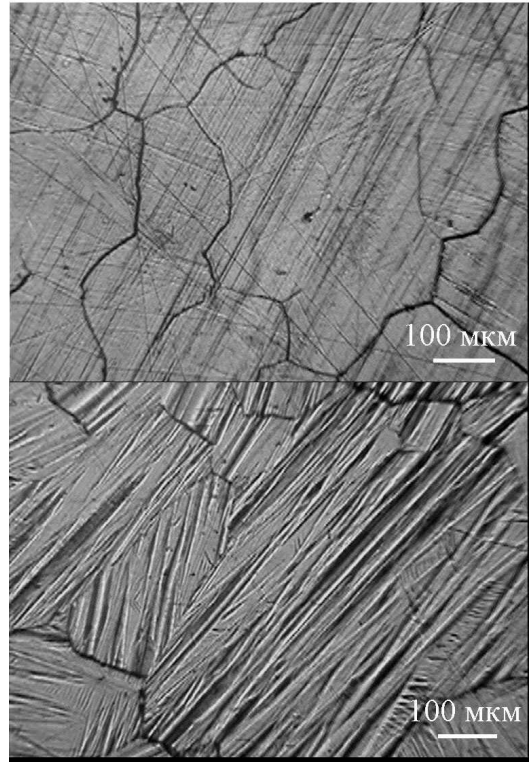}
\caption{RT optical image of the Fe$_{45.2}$Mn$_{25.9}$Ga$_{28.9}$
alloy sample surface subjected to heating up to $T$=373 K (upper
panel) and cooling down to $T$=78 K (bottom panel).} \label{fig6}
\end{figure}

Conclusions on the sample structures based on the results of XRD
measurements are supported also by RT optical image of the sample
surface - martensitic variants dominate on the surface of the
sample previously cooled down the liquid nitrogen temperature (see
Fig. \ref{fig6}).

The obtained lattice parameters for cubic austenite and
martensitic tetragonal phases nicely agree with the results
obtained in the literature. Zhu \emph{et al.} the marentsitic
phase for Fe$_{50}$Mn$_{22.5}$Ga$_{27.5}$ HA at $T$ = 90 K have
been indexed as a body-centered tetragonal structure with the
lattice parameters $a=b$ = 0.5328 nm and $c$ = 0.7113 nm and
largest lattice distortion, $(c-a)/a=33.5 {\%}$ among all the
reported FSMAs \cite{zhu}. The similar result has been obtained by
Omori \emph{et al.} for Fe$_{44}$Mn$_{28}$Ga$_{28}$ HA -
martensitic phase has been identified as non-modulated tetragonal
one with the lattice parameters of $a=b$ = 0.5368 nm, $c$ = 0.7081
nm and the $c/a$ = 1.319, where the distorted L2$_{1}$ structure
is taken as the unit cell of the non-modulated tetragonal
martensite. The volume change due to the forward transformation
$\Delta$V/V is +0.73 $\%$ \cite{omori1}.

In our case lattice distortion $(c-a)/a$ of the tetragonal phase
is 34.5 ${\%}$. Forward MT induces the volume increase of
$\Delta$V/V = 3.28 $\%$.

Abnormally intense (400) reflection for L2$_{1}$ phase indicates
on the presence of some texture in polycrystalline
Fe$_{45.2}$Mn$_{25.9}$Ga$_{28.9}$ alloy sample.

\subsection{Magnetic properties}
\label{5}

\begin{figure}[t]
\includegraphics[width=8.5cm]{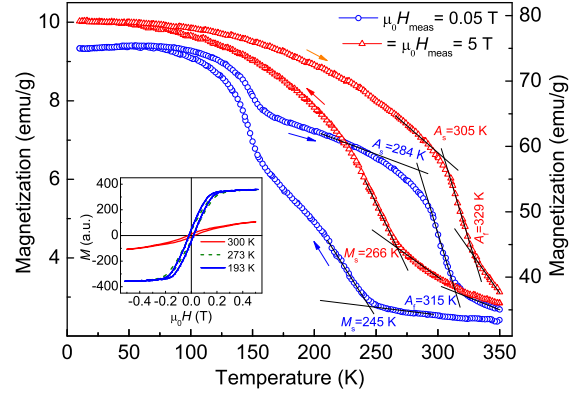}\caption{Temperature
dependencies of the magnetization of
Fe$_{45.2}$Mn$_{25.9}$Ga$_{28.9}$ polycrystalline alloy measured
in a weak ($H$ = 500 Oe, left scale) and strong ($H$=50000 Oe,
right scale) magnetic fields on sample heating and cooling. Inset
shows magnetization hysteresis loops of alloy taken at different
temperatures upon sample cooling.} \label{fig7}
\end{figure}

\begin{figure}[t]
\includegraphics[width=8.5cm]{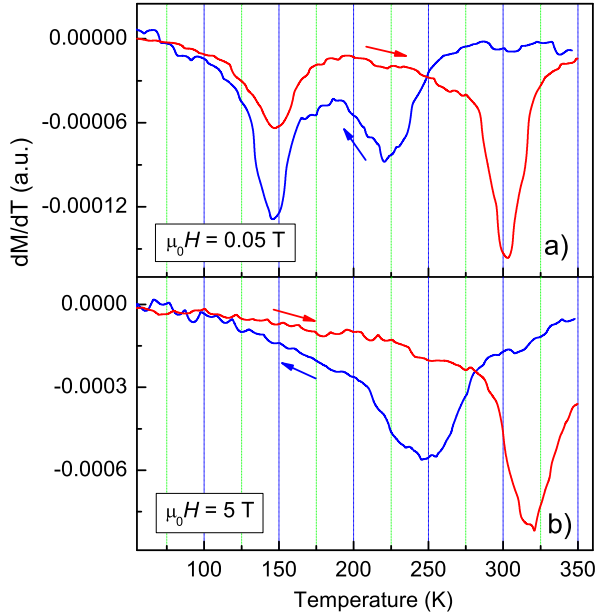}\caption{Temperature
derivatives of $M(T)$ plots for Fe$_{45.2}$Mn$_{25.9}$Ga$_{28.9}$
HA obtained at different measuring fields on sample heating and
cooling.} \label{fig8}
\end{figure}

Magnetic properties of bulk Fe$_{45.2}$Mn$_{25.9}$Ga$_{28.9}$
polycrystalline alloy sample are presented on Figs. \ref{fig7} and
\ref{fig8}. The temperature dependence of magnetization $M(T)$
obtained at weak ($H=$ 500 Oe) magnetic field exhibits two
definite drops upon sample heating from $T=$ 4 K. Reverse sample
cooling from $T=$ 350 K causes an increase of magnetic moment also
at two steps started at $T\approx$ 248 and 153 K. It is clearly
seen that the low-temperature peculiarity does not depend on
heat-treatment direction while the high temperature one does (see
Figs. \ref{fig7} and \ref{fig8}). Such a behavior of $M(T)$ and
$dM/dT$ dependencies supports two-phase nature of the sample
revealed by XRD data: low-temperature peculiarity at $T\approx$
150 K can be definitely ascribe to the Curie temperature of the
cubic L2$_{1}$ phase while high-temperature one is concerned with
MT between tetragonal and cubic phases.

It can be concluded that the Curie temperature for tetragonally
distorted L2$_{1}$ phase at least, not less than $T\approx$ 315 K.
Thus, for 150 $\leq T \leq {329}$ K temperature range we
have the deal with nonmagnetic (i. e. PM) L2$_{1}$ phase and FM
tetragonal matrensitic one.

Increase in the external magnetic field up to 50 kOe drastically
changes the $M(T)$ and $dM/dT$ dependencies for
Fe$_{45.2}$Mn$_{25.9}$Ga$_{28.9}$ alloy: low-temperature
peculiarities concerned with the Curie temperature of L2$_{1}$
phase disappear, {$A_{f}$ temperature increases up to about
329 K, the hysteresis between $M(T)$ plots obtained on the sample
heating and cooling and determined as $A_{f}$ - $M_{s}$ converges
from ${\Delta}T_{H = 500 Oe}{\approx}$ 70 K down to ${\Delta}T_{H
= 5 kOe}{\approx}$ 63 K (see Figs. \ref{fig7} and \ref{fig8}).} It
should be noted here that the saturation magnetization of
Fe$_{45.2}$Mn$_{25.9}$Ga$_{28.9}$ alloy at $T=$ 4 K and
${\mu}_0{}H$=5 T $M_{sat}$=79 emu/g (or 3.26 $\mu_{B}$/f.u.) is
close to those reported in the literature: 71 emu/g for
Fe$_{44}$Mn$_{28}$Ga$_{28}$ alloy, 83 emu/g for
Fe$_{43}$Mn$_{28}$Ga$_{29}$ alloy, 93.8 emu/g for
Fe$_{50.0}$Mn$_{22.5}$Ga$_{27.5}$ alloy \cite{zhu,omori1,omori2}.
Recall that calculated magnetic moments for L2$_{1}$ and L1$_{0}$
phases are ${\mu}=$ 2.01 and 6.35 $\mu_{B}/f.u.$, respectively.
Therefore, taking into account two-phase nature of
Fe$_{45.2}$Mn$_{25.9}$Ga$_{28.9}$ alloy, experimentally determined
its saturation magnetization value looks reasonable.

According to the results of first-principle calculations the
specific magnetic moments of the L2$_{1}$ and tetragonal phases
differ more than three times. Concerning $M(T)$ dependence
obtained at $H=$ 500 Oe field (see Fig. \ref{fig7}) the amount of
L2$_{1}$ phase in Fe$_{45.2}$Mn$_{25.9}$Ga$_{28.9}$ alloy at
$T\approx$ 150 K is comparable with tetragonal one. Therefore,
almost invisibleness of FM to PM transition related to L2$_{1}$
phase in the $M_{H=50kOe}(T)$ plot can be explained by complete
L2$_{1}$ to tetragonal phase transition induced by strong magnetic
field rather than smearing in temperature of FM to PM transition
in L2$_{1}$ phase.

{The high-temperature shift of the MT temperatures with
magnetic field $H$ once again supports an assumption that the
Curie temperature for a tetragonal phase is not less than
$T\approx$ 329 K.}

It is known that the magnetic field is the factor which can play
an important role in thermodynamical processes. {For
example, the effect of a magnetic field on the temperatures of
direct and reverse martensitic transformations was investigated
for Ni$_{2+x}$Mn$_{1-x}$Ga alloys in a pioneering paper of
Dikstein \emph{et al}. \cite{dik}.} The shift of the phase
equilibrium temperature under the influence of magnetic field can
be expressed by Krivoglaz-Sadovsky equation \cite{krivo}:

\begin{equation}
\Delta T =T_{0}(m_{1}V_{1}-m_{2}V_{2}){\Delta}H/q ,
\end{equation}

where $T_{0}$ is the temperature of phase transition without
magnetic field, $m_{1,2}$ and $V_{1,2}$ are magnetic moments and
volumes of initial (1) and final (2) phases, ${\Delta}H$ is the
change of external magnetic field, $q$ is the specific heat of
transition. For the case of reverse MT in
Fe$_{45.2}$Mn$_{25.9}$Ga$_{28.9}$ alloy transition from FM
martensitic phase to PM austenite L2$_{1}$ phase ($m_{2} =
m_{L21}$${\approx}$ 0 for 150 ${\leq}T{\leq}$ 312 K temperature
range) this equation can be written as:

\begin{equation}
\Delta T =T_{0}m_{1}V_{1}{\Delta}H/q ,
\end{equation}

It is clear that for our case external magnetic field should
induce a positive shift of phase transition temperature. Indeed,
according to the results shown in Figs. \ref{fig7} and \ref{fig8}
this shift can be evaluated as {${\Delta}T$ = $A_{f}(H=50
kOe)$ - $A_{f}(H=500 Oe){\approx}$ 14 K for reverse MT or
${\Delta}T$ = $M_{s}(H=50 kOe)$ - $M_{s}(H=500 Oe) {\approx}$ 21 K
for direct MT.}

Magnetization hysteresis loops for
Fe$_{45.2}$Mn$_{25.9}$Ga$_{28.9}$ polycrystalline alloy sample
indicate on its rather high coercivity $H_{c}\approx$ 200 Oe and
magnetization saturation field $H_{s}\approx$ 2000 Oe (see inset
in Fig. \ref{fig7}).

\subsection{Transport properties}
\label{6}

\begin{figure}[t]
\includegraphics[width=8.5cm]{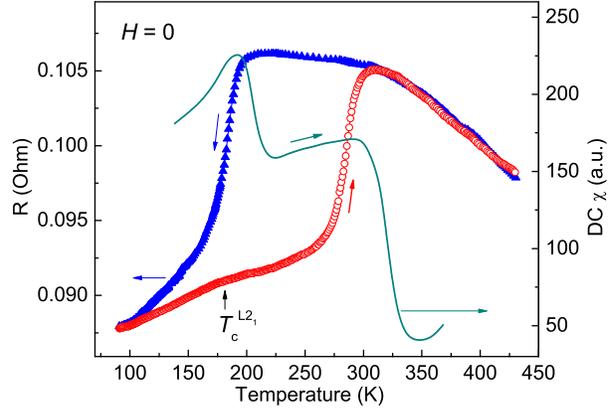}
\caption{Temperature dependencies of resistance for bulk
Fe$_{45.2}$Mn$_{25.9}$Ga$_{28.9}$ HA obtained on sample heating
and cooling at zero magnetic field (symbols, left scale) and DC
magnetic susceptibility obtained at $H_{meas}\approx$ 5 Oe upon
sample heating (line, right scale).}\label{fig9}
\end{figure}

Figure \ref{fig9} shows the temperature dependencies of resistance
$R(T)$ and DC magnetic susceptibility ${\chi}(T)$ for bulk
Fe$_{45.2}$Mn$_{25.9}$Ga$_{28.9}$ alloy obtained at very weak
measuring magnetic field. Increase in temperature from $T=$ 78 K
causes a nearly linear growth of resistance with the positive
temperature coefficient of resistivity (TCR).  At $T\approx$ 175 K
small break of the $R(T)$ plot can be seen, TCR value becomes a
little bit smaller. At $T\approx$ 272 K resistance rapidly growths
by about 13 $\%$ indicating start of reverse MT in an alloy. At
$T\approx$ 299 K reverse MT is finished and resistance of alloy
starts to decrease almost linearly with temperature. It should be
noted here that the temperatures of the peculiarities on the
${\chi}(T)$ plot coincide with those on the $R(T)$ plot obtained
on the sample heating. Taking into account the results of magnetic
measurements (see Figs. \ref{fig7} and \ref{fig8}) low-temperature
break of the TCR value can be definitely ascribe to the FM to PM
transition in L2$_{1}$ phase of an alloy. Indeed, electron-magnon
(spin-disorder) scattering usually reaches its maximum near the
Curie temperature; above $T_{C}$ the spin-disorder mechanism is
independent of $T$. Therefore, for some ferromagnetic metals and
alloys (including FM HA), a distinct change in the slope of the
$\rho(T)$ dependence can be expected at $T_{C}$
\cite{majumdar,bose,kudrnovsky,barth}.

Upon sample cooling from $T=$ 425 K resistance follows the same as
on the heating way up to $T\approx$ 299 K where some break of the
negative TCR value can be seen. The rapid drop of resistance at
$T\approx$ 195 K indicates on the start of the forward MT which is
finished at $T\approx$ 169 K. Thus, MT in
Fe$_{45.2}$Mn$_{25.9}$Ga$_{28.9}$ alloy causes the significant
changes of the resistance value and the change of the TCR sign.
Such a behavior is the direct consequence of the differences in
the electronic structures of these phases, namely DOS values at
$E_{F}$ for L2$_{1}$ and tetragonal phases of Fe$_{2}$MnGa alloy
(see Fig. \ref{fig1}).

It should be mentioned here that the calculated value of
resistivity of Fe$_{45.2}$Mn$_{25.9}$Ga$_{28.9}$ alloy sample
shows unreasonably high-value $\rho_{78K}=$3.08 m$\Omega$cm
probably due to internal cracks induced by sample cutting and
concerned with sample brittleness. Therefore, the values of the
resistivity Fe$_{2}$MnGa alloys should be treated with caution.

Temperature dependencies of resistivity for bulk
Fe$_{45.2}$Mn$_{25.9}$Ga$_{28.9}$ HA obtained on sample heating
and cooling exhibit huge temperature hysteresis - ${\Delta}T_{R}
\approx$100 K. This value is rather close to hysteresis of
magnetization obtained at $H=$ 500 Oe - ${\Delta}T_{M}\approx$ 90
K. Such a huge value of thermal hysteresis of structure and
physical properties is probably concerned with great friction
between the boundaries of martensite and austenite phases due to
large lattice distortion at MT.

\subsection{Optical properties}
\label{7}

\begin{figure}[h]
\includegraphics[width=8.5cm]{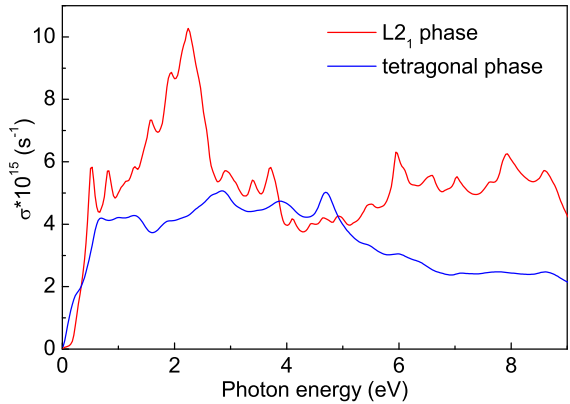}
\caption{Calculated interband optical conductivity spectra of
stoichiometric Fe$_{2}$MnGa alloy in cubic L2$_{1}$ and tetragonal
phases.}\label{fig10}
\end{figure}

Figure \ref{fig10} presents calculated interband optical
conductivity spectra for perfectly ordered stoichiometric
Fe$_{2}$MnGa alloy with L2$_{1}$ and L1$_{0}$ types of atomic
order. Both spectra are characterized by the set of interband
transitions which form interband absorption peak at 0 $\leq
\hbar\omega\leq$ 5 eV energy range. However, this peak for
L2$_{1}$ phase has a larger intensity and is manifested more
definitely. This result looks as unexpected considering noticeable
difference in energy dependencies of the DOS values for these
phases (see Fig. \ref{fig1}).

\begin{figure}[h]
\includegraphics[width=8.5cm]{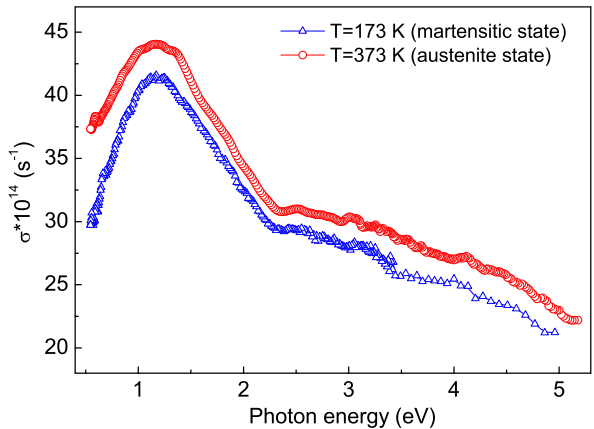}
\caption{Experimental optical conductivity spectra of bulk
Fe$_{45.2}$Mn$_{25.9}$Ga$_{28.9}$ HA in austenite ($T_{meas}$ =
373 K, circles) and martensitic ($T_{meas}$ = 173 K, triangles)
states.}\label{fig11}
\end{figure}

Experimental optical conductivity spectra of bulk
Fe$_{45.2}$Mn$_{25.9}$Ga$_{28.9}$ HA at different temperatures (i.
e. in different structural states) are shown on Fig. \ref{fig11}.
In order to obtain confidently the austenitic and martensitic
states in the alloy, the optical measurements were carried out at
$T$ = 373 K and $T$ = 173 K on the sample preliminary heated and
cooled up to 573 and down to 78 K, respectively. Such an approach
allows us definitely to fix martensitic and austenite phases in
alloy and to minimize the temperature effect on the optical
properties of the alloy. Unlike theoretical predictions the
optical conductivity spectra for austenitic and martensitic phases
look rather similar and both are characterized by broad intense
interband absorption band located at $\hbar\omega\approx$ 1.2 eV.

To the best of our knowledge, there are few publications devoted
to the experimental study of the optical properties of
Fe$_{2}$MnGa alloys \cite{kudr2,kral}. Thus, Kr$\acute{a}$l
investigated the optical properties of several electrolytically
polished bulk Fe-Mn-Ga alloys near the stoichiometry 2:1:1. All
the investigated by him alloys of different compositions (and
probably different crystalline structures) demonstrate rather
similar optical properties \cite{kral}. Furthermore, the optical
conductivity spectrum for bulk Fe$_{46.6}$Mn$_{24.2}$Ga$_{29.2}$
HA obtained by Kr$\acute{a}$l in the spectral shape and absolute
value practically coincides with that for
Fe$_{45.2}$Mn$_{25.9}$Ga$_{28.9}$ HA studied in a present work and
shown on Fig. \ref{fig11}.

Such a similarity of the experimental optical properties of bulk
Fe$_{45.2}$Mn$_{25.9}$Ga$_{28.9}$ HA in the martensitic and
austenitic phases probably can be concerned party with two-phase
nature this alloy sample, whose optical properties predicted to be
rather similar. At the same time the two-phase nature of
Fe$_{45.2}$Mn$_{25.9}$Ga$_{28.9}$ alloy sample was not sufficient
for a complete masking of the changes in its magnetic and
transport properties induced by MT because of the properties of
pure austenitic and martensitic phases differ drastically.

\section{Summary}
\label{8}

1. First-principle calculations revealed that the stoichiometric
Fe$_{2}$MnGa alloy with L2$_{1}$ and L1$_{0}$ types of the atomic
order has noticeable different $N(E)$ dependencies (especially at
$E_{F}$), exhibits three times different magnetic moments of the
formula units and manifests qualitatively similar calculated
interband optical conductivity spectra.

2. Temperature-induced direct martensitic transformation in
Fe$_{45.2}$Mn$_{25.9}$Ga$_{28.9}$ Heusler alloy causes a rapid
decrease in its resistivity value, changes of the TCR sign from
negative to positive and leads to the transition from PM to FM
state.

3. Direct and reverse MT in Fe$_{45.2}$Mn$_{25.9}$Ga$_{28.9}$
Heusler alloy demonstrate about $\Delta$$T$$\approx$ 90 - 100 K
hysteresis in critical temperatures manifested by temperature
dependencies of resistance and magnetization.

4. External magnetic field of $H$ = 50 kOe causes the positive
shift of the critical temperatures and reduces temperature
hysteresis width.

5. Experimentally observed changes in the magnetic and transport
properties of Fe$_{45.2}$Mn$_{25.9}$Ga$_{28.9}$ alloy induced by
MT nicely can be explained by the changes in the electronic
structures of alloy upon transition from the austenitic to
martensitic phase.

6. MT does not lead to visible changes in the experimental OC
spectra of Fe$_{45.2}$Mn$_{25.9}$Ga$_{28.9}$ alloy because of
partly its two-phase nature and not very significant difference in
calculated optical properties of these phases.

\section{Acknowledgments}

This work has been supported by the project ``Marie
Sk{\l}odowska-Curie Research and Innovation Staff Exchange
(RISE)'' Contract No. 644348 with the European Commission, as part
of the Horizon2020 Programme. We are also grateful to V. K.
Nosenko and E. O. Svistunov for assistance and critical
discussions.

\end{document}